\documentclass[quaderni]{sifrev}        

\usepackage{url}

\usepackage{endnotes}

\begin{document}

\title{Enrico Fermi and the Old Quantum Physics}

\author{Alberto De Gregorio \and Fabio Sebastiani}  


\institute{Dipartimento di Fisica, ``Sapienza'' Universit\`{a} di Roma\\ \email{alberto.degregorio@roma1.infn.it \\  fabio.sebastiani@roma1.infn.it}}     

\vol{}                                  
\issue{}				  

\month{}                        


\maketitle

\begin{abstract}

{\bf ABSTRACT --} We outline Fermi's attitude towards old quantum physics, which he studied autonomously, and we focus on his resort to old quantum physics to work out quantum statistics and teach atomic physics. In Section 1, we sketch out the context from which Fermi's interest for quantum physics arose. In  Section 2,  we deal with Fermi's learning and early production in quantum physics. In  Section 3, we go through Fermi's work on quantum statistics. As further  evidence for our reconstruction, we also go through the first two courses on theoretical physics he held in Rome: we deal with Fermi's 1926-27 course in  Section 4, while in  Section 5 we discuss his 1927-28 course and his 1928 book on atomic physics.
\end{abstract}

\onecolumn
 
\section{The context}         
\label{sec:foo}

In winter 1925-26, Fermi worked out the quantum statistics named after him. It represented the last relevant result obtained in the framework of the old quantum physics.  
Fermi had a clear preference for a phenomenological approach, as we can conclude also from analysing his early teaching activity. This predilection fitted in with the semiclassical theory; on the other hand, it might well be that his autonomous study of the old quantum physics, in winter 1918-19, stirred his very preference for a phenomenological approach. 

Until mid 1920s, atomic theory relied on the so called old quantum physics, which originated from the pioneering works of Max Planck, Albert Einstein, and Niels Bohr and was further developed by Arnold Sommerfeld, Paul Ehrenfest, as well as Bohr and Einstein themselves. The cornerstones of the old quantum physics were:\footnote{ 
See \cite{Jammer}, \cite{Tagliaferri}, \cite{MehraRechenberg}.}

\begin{itemize}
\item the introduction of the energy element $\epsilon= h\nu$  (Planck, 1900);

\item the quantum condition for the angular momentum of atomic electrons, $M=nh/2\pi$, which generalised the Planck-Einstein condition for the energy (Bohr, 1913);

\item the frequency condition $\nu=(E-E^\prime)/h$, implying that the absorption and emission of electromagnetic energy involve the transition between two stationary states, of energy $E$ and $E^\prime$ respectively (Bohr, 1913);

\item the correspondence principle, which prescribed that the results of quantum physics agree with the classical ones, in the limit of large quantum numbers (Bohr, 1913);

\item the quantum conditions $\oint{p_idq_i=n_ih}$ for multiperiodic motions, a generalisation of the quantum condition for the angular momentum (Sommerfeld, 1915-16);

\item the adiabatic hypothesis, which provides that the quantum state of a system does not vary under infinitely slow (``adiabatic'') changes of the potential (the Sommerfeld quantum conditions are adiabatically invariant) (Ehrenfest, 1916-17).\footnote{
A most plain and comprehensive exposition of the Ehrenfest adiabatic principle was 
\cite{Levi-Civita}). 
Tagliaferri recalled that the Sommerfeld conditions were already well accepted when Ehrenfest worked out the adiabatic principle, which therefore turned out more relevant for conceptual than for practical purposes (see \cite{Tagliaferri}, p. 219).}
\end{itemize}

The new quantum mechanics that superseded the old quantum theory was pioneered by Heisenberg (summer 1925) and Schr\"odinger  (winter 1925-26). It soon became a fully comprehensive theory, accounting for many properties of the atom.

In those same years, physics research in Italy was under the banner of an exaggerated empiricism, and turned out too much wary and slow in following the new course. There certainly were some well versed physicists; though, their duty appeared to provide only for accurate measures of (often marginal) phenomena.\footnote{
See 
\cite{GiulianiMarazzini}, \cite{Giuliani}, \cite{Sebastiani}.}
In the very meanwhile, the theoretical foundations of physics were being worked out abroad.

Enrico Persico wrote that ``if we were to leaf through the Nuovo Cimento dating around 1920, we would see that most of the physics published in Italy dealt with intricate and obscure matters which could only be studied empirically without a clear theoretical guide [...]. Many of these phenomena in the light of today's experience can be more or less laboriously interpreted, but they appear to be a snarl of many different elementary phenomena and thus, particularly unsuitable as the key to the really fundamental questions''.\footnote{
\cite{Persico}, pp. 42-43 
(see also \cite{PersicoSouvenir}, pp. 322-23).}

There was no true school of theoretical physics in Italy until the second half of 1920s, as most of Italian physicists ignored the tumultuous progress abroad.\footnote{
Italy excelled, and was prominent in Europe, for what concerned mathematical physics. The latter aimed mainly at the resolution of the mathematical issues arising from physical questions.}
The International Conference held in Como (and in Pavia and Rome) in September 1927 marked the turn of the Italian physics towards the new theories.

At that Conference, Bohr gave a very preliminary presentation of the principle on the complementary aspects of quantum physics.\footnote{
\cite{BohrComo}.
About Bohr's talk in Como, see 
\cite{DG-S}.}
Urged also by Sommerfeld's talk, much attention in the discussions was paid to the Fermi statistics. That was the first relevant result obtained in quantum physics by an Italian, and was still based on old quantum theory.\footnote{
In an interview with Thomas Kuhn, Rasetti said that the Como Conference had an enormous effect on the development of Italian physics, because it ``was the revelation of Fermi to the Italians.'' (\cite{PersicoRasetti-Kuhn}, p. 2/2)}

\section{Fermi's learning and early production in quantum physics}
\label{sec:foo}

Fermi was essentially a self-taught physicist.\footnote{
See \cite{SassiSebastiani}, \cite{PetrucciSebastiani}.
}
In his youth (he was born in 1901) he became persuaded that a too rigorous approach should not prevail over a phenomenological one; formal rigour might rather be regained later. This belief marked his research activity, as well as his teaching style. In fact, Emilio Segr\`e held that Fermi's ``clear, simple, easy'' way of teaching ``was often the result of deep meditation on his part, started in early youth, when he was his own pupil''.\footnote{
\cite{SegrePreface}, p. v.}

When Fermi entered the {\it Scuola normale superiore} in Pisa, in autumn 1918, he already knew many of the topics taught at the university. He took his first steps towards quantum physics autonomously during the first semester of 1919, going through Planck's {\it Vorlesungen \"uber die Theorie der W\"armestrahlung} (Lectures on the Theory of Heat Radiation) and Owen W. Richardson's treatise {\it Electron theory of matter}.\footnote{
See 
\cite{SassiSebastiani}. The chapter of Richardson's {\it Electron theory of matter} devoted to the structure of the atom deals almost exclusively with Bohr's 1913 celebrated trilogy.}
Persico remarked that ``even though [Fermi] diligently attended the courses and sessions, his studies focused on subjects which he had personally chosen independently of his obligatory curriculum. For example, in February 1919, that is during his first year at the University, he wrote to me: `... since I have almost nothing to do for school and since there are lots of books available here, I am trying to learn more about mathematical physics and I will try to do the same for mathematics'.''\footnote{
\cite{Persico}, p. 37.}

Starting from the end of 1919, Fermi considered that he already mastered the works anterior to 1915, and went through the most relevant results published afterwards (in particular, he studied {\it Atombau und Spektrallinien}, which Sommerfeld published just in 1919).
His autonomous study was absolutely fruitful. Franco Rasetti, Fermi's fellow student, recalled that ``such topics as the quantum theory had gained no foothold in Italy: they represented a `no man's land' between physics and mathematics. Fermi was the first in the country to fill the gap''.\footnote{
\cite{RasettiForeword}, p. 56.}
Further, ``by 1920 or even '22, quantum theory in Italy was essentially confined in Fermi's mind and there was very little outside''.\footnote{
\cite{PersicoRasetti-Kuhn}, p. 8/1.
Actually, Antonio Garbasso had already applied the Bohr model in Italy before Fermi, in 1914
(\cite{Garbasso1},
\cite{Garbasso2}). For Garbasso's warm welcome to Bohr's theory, see
\cite{Brunetti}, and
\cite{Leone}.}

On January 30, 1920, Fermi wrote to Persico: ``I am gradually becoming most influential at the Physics Institute. To be sure, one of the next days I shall give a conference on quantum theory before many magnates. I really am a propagandist of quantum theory, particularly for what concerns spectral phenomena''.\footnote{
Letter reported in \cite{AnniItaliani}, p. 83.}
In particular, Fermi's wife, Laura Capon, recalled that Fermi and Rasetti ``knew much more physics than [their professor of experimental physics Luigi Puccianti]. He was aware of this fact, and he asked Fermi to teach him theoretical physics''.\footnote{
\cite{Capon}, p. 25.}

In May 1920, Fermi was involved in a problem of theoretical spectroscopy: he definitely proved to have full command of the Bohr-Sommerfeld model.\footnote{
See \cite{AnniItaliani}, pp. 82-86.}

Despite being tied on an experimental work for his thesis (on X-rays),\footnote{
In 1962,
G. Polvani bemoaned the failure of numerous attempts to trace Fermi's thesis (see \cite{FNM}, vol. I, p. 35). The manuscript of Fermi's thesis was recovered by Roberto Vergara Caffarelli at the Pisa University only a few years ago: its spine reported the name ``Terni,'' instead of ``Fermi.''}

Fermi wrote a long preface, dealing both with the Bohr-Sommerfeld theory of the atom and the Einstein-Debye theory of solids. A part of Fermi's thesis, accounting extensively for, and applying, the Bohr-Sommerfeld model, was published in {\it Il Nuovo Cimento}.\footnote{
See
\cite{FermiRoentgen}.
Fermi investigated the characteristic lines, using the Bohr-Sommerfeld model. The second chapter of his thesis was devoted to the dependence of the diffraction patterns on the temperature, which he tackled --- following a suggestion of Debye himself --- in the framework of the Einstein-Debye theory. Fermi wrote two unpublished manuscripts (\cite {Unpublished1}, \cite {Unpublished2}) out of  this second chapter.
The two manuscripts
are kept in Pisa (see
\cite{Leone2}, pp. 517-18).}
Fermi's approach was very pragmatic, disclosing his special preference for phenomenology. In his thesis, he warned the reader that a given hypothesis should not lead to ``too much intricate calculations, unless being justified by the final result''.\footnote{
\cite{Tesi}, p. 41.}
In fact, he charged that some of his theoretical physics colleagues behaved like devotee of vain formalism.

After his degree, Fermi came back to Rome and was introduced to the Director of the Physics Institute of via Panisperna, Orso M. Corbino, who immediately realised his extraordinary talent and provided for his completing his education in G\"ottingen, with Max Born, and in Leiden, with Ehrenfest, in 1923 and 1924 respectively.\footnote{
See 
\cite{CSGottinga},
\cite{CSStatistics}.}

In G\"ottingen, Fermi found some of ``the brightest luminaries of theoretical physics,'' among others the young Werner Heisenberg and Pascual Jordan. However ``he remained aloof'' from the physicists of his age.\footnote{
\cite{EFPhysicist}, pp. 32-33.}

Born had a thorough formal approach to atomic physics, which Fermi did not really like\footnote{
See \cite{CSStatistics}.}
--- even if he became acquainted in G\"ottingen with both the subtleties of the perturbative theory of the helium atom, and the quantisation of the ideal gas, which would enable him to work out the new quantum statistics. 

In an interview with Kuhn, Maria G. Mayer recalled that ``Born was really very mathematical [...] never looking much at the physics.'' As for the atmosphere in G\"ottingen, Mayer added that ``Ehrenfest used to come every summer to G\"ottingen, and Ehrenfest taught us the physics. [...] He was a wonderful complement to Born''.\footnote{
\cite{Mayer}, p. 3.}

In 1924, Fermi found ``a very congenial atmosphere'' in Leiden, where he became friend with many young physicists, and ``Ehrenfest encouraged him greatly''.\footnote{
\cite{EFPhysicist}, p. 36.}
He also met with Einstein who, during a short stay in Leiden, was extending the quantum statistics (that Bose had worked out for light quanta) to ideal gases. 

While in Florence --- where he taught mathematical physics for two years --- in 1925 Fermi published {\it Sui principi della teoria dei quanti} (On the principles of quantum theory), his first popular writing about quantum theory.\footnote{
\cite{FermiDivulg}.
See also \cite{Petrucci}.}
Two fundamental principles are stated at the beginning: (a) ``[...] Not all the motions that are mechanically allowed do really take place, but only those belonging to a discrete sequence of quantum, or static [i.e. stationary], states.'' As a consequence, the energy $w$ of the system takes a discrete sequence of values, the so-called energy levels. (b) ``Radiation-energy emission always depends on the non-mechanical electron jump between two quantum states of motion:'' the energy is carried by ``one quantum only,'' of frequency $\nu= (w-w')/h$.\footnote{
\cite{FermiDivulg}, p. 139.}

These two principles had been tested experimentally in many ways, but they failed ``to solve the whole of problems of atomic physics.'' Fermi referred to the Ehrenfest adiabatic principle and to the Bohr correspondence principle, which helped solve, ``in part at least,'' some of these issues. Fermi illustrated the Ehrenfest principle in the case of a harmonic oscillator that changes its ``mechanism'' (i.e. its elastic constant $k$) very slowly (adiabatically): the ratio between its energy and its frequency is adiabatically invariant, and ``if the relation $w/\nu=nh$ held before the transformation, it holds also during the whole transformation,'' so that the oscillator remains in the same quantum state as before. The same property held for any mechanical system. 

Fermi then did not follow the historical development of the theory and showed how to choose from all possible mechanical motions, following the adiabatic principle: ``The phase integrals  $J_i=\int{p_idq_i}$ are adiabatic invariant along a whole oscillation [...]. According to the Ehrenfest principle, the $J$'s clearly are the quantities to be quantised.'' For the oscillator, $J=w/\nu$ leads to the quantum condition $J=nh$. The latter may be generalised, and leds to the Bohr-Sommerfeld conditions $J_i=n_ih_i$ for systems with separable variables. 

As for Bohr correspondence principle, Fermi stressed how deeply it differs from the other principles of mathematics and physics, since it does not provide any definite statement like $A=B$ or $A>B$, but only statements like ``{\it A} is similar to {\it B}.'' For instance, ``the intensity and polarisation of the lines, an atom emits in conformity with the quantum theory, resemble the intensity and polarisation of the lines provided for by the classical theory. [...] The strict identity holds in the limit [of high quantum numbers] only; of course, for finite quantum numbers, a mere similarity corresponds to this identity.''

Fermi wrote at the end: ``We can then conclude that quantum theory still fails in giving solution to some fundamental problems, but any attempts to logically deduce all its rules from general principles are being made.''

Fermi's paper provided a concise and very clear description of the old quantum physics, at the very same time when Heisenberg, Born, and Jordan were conceiving matrix mechanics in G\"ottingen. 

Fermi published many papers on the quantum theory, also in prestigious German scientific journals, between 1922 and 1925. They were a real breakthrough in the Italian outline, steering and interesting the youngest people to the new physics. Fermi's wide and distinguished scientific activity of these years culminated in the formulation of the quantum statistics named after him, at the beginning of 1926.\footnote{
See \cite{CSStatistica}.}

\section{Quantum statistics}
\label{sec:foo}

Fermi's work on the quantum statistics, {\it Sulla quantizzazione del gas perfetto monoatomico} (On the quantisation of the monoatomic ideal gas), was communicated for the first time by Antonio Garbasso, before the {\it Accademia dei Lincei}, on February 7, 1926.\footnote{
\cite{FermiPerfetto}.}
On March 26, the editor of {\it Zeitschrift f\"ur Physik} received a paper with a more detailed exposition, eventually published on May 11. The aim and the results of that work were outlined in the Summary as follows: 

\begin{quote}
All theories dealing with the degeneration [of the ideal gas at low temperature] make more or less arbitrary assumptions about the statistical behaviour of molecules, or about their quantisation. It shall in the present work be assumed only the rule, that was proposed for the first time by Pauli and proved by many spectroscopic facts, that two equivalent elements can never exist in a system, whose quantum numbers completely agree. The equation of state and internal energy of the ideal gas shall be deduced from this assumption; the entropy at high temperature agrees with the results of Stern and Tetrode.\footnote{
\cite{FermiIdealen}, p. 186. For an account of the birth of the Fermi statistics, see \cite{RasettiPrefazione}, \cite{AmaldiStatistics}.}
\end{quote}

Fermi, as well as all the physicists engaged in the quantum-statistical treatment of systems of identical particles, regarded the molecules as indistinguishable. He quantised the gas by supplementing the Sommerfeld conditions with a new rule, determined by the exclusion principle: ``Since this rule of Pauli has proved extremely fruitful in the explanation of spectroscopic facts, we propose to inquire whether it is not also of some use for the problem of the quantisation of ideal gases.'' This ``addition to the Sommerfeld quantum conditions'' became a keystone, providing Fermi with ``a method of quantization of an ideal gas which, in our opinion, is as independent of arbitrary assumptions about the statistical behaviour of the gas molecules as is possible.''\footnote{
\cite{FermiIdealen}, {\it passim}.}
Fermi's method appeared as an extension, rather than a mere application, of the exclusion principle. 

The motions of the molecules in a gas needed to be quantised, in order to provide the quantum numbers to which the Pauli principle applied. The quantisation of a gas in a box, introduced by Brody (an assistant of Born in G\"ottingen) in 1921, was the simplest and most immediate way to do it.\footnote{
\cite{Brody}. Brody applied the Sommerfeld quantum conditions to the gas molecules in a box: his methods led to the same results as quantum mechanics.}

Though Fermi was well acquainted with Brody's method of attack, he chose another approach --- which he in the German paper stated to be ``more general'' ({\it allgemeiner}), and also ``more convenient for the calculations'' in the Italian paper: he imposed a central elastic force on the molecules, directed towards a fixed point, so that each molecule became a tridimensional harmonic oscillator; the final result did not depend on the special choice of the force field. 

Rasetti wrote: ``It is noteworthy that Fermi did not enclose his ideal gas in a box, according to the current treatment, but placed the particles in a three-dimensional harmonic oscillator potential. In this way he obtained a spherically symmetric, monotonically decreasing gas density. For large radii, the density was always sufficiently low to ensure validity of the classical Boltzmann approximation and hence to define the temperature for the entire gas''.\footnote{
\cite{RasettiPrefazione}, p. 178.}
However, Fermi would not use this model anymore, choosing instead the usual box in his works on quantum statistics to come.

The harmonic potential allowed Fermi to use the Ehrenfest adiabatic principle, which he had dealt with three years before (and also in 1925, in his popular writing on the old quantum physics).\footnote{
\cite{FermiConsiderazioni}, \cite{FermiAdiabatiche}.}
Actually, the quantum number $s$ of an harmonic oscillator of energy $E=sh\nu$ did not change under an adiabatic variation of its elastic constant, in accordance with Ehrenfest's adiabatic principle. The same could not be said about the quantum numbers of the molecules in a box, under an adiabatic displacement of its walls. 

Fermi formulated a simple and subtle definition of the gas temperature, independent of more or less arbitrary assumptions about the probability of a state (assumptions unavoidable when using the relations $T=dW/dS$ and $S=k\ln P$, where $T$ is the temperature, $W$ the energy, $S$ the entropy, and $P$ the number of microstates). Actually, for infinitely large $r$ the gas density vanished and the degeneration phenomena disappeared, so that the statistics of the gas reduced to the classical one (in particular, for $r\rightarrow\infty$ the velocity distribution converged to the Maxwellian one). But the entire gas was at the same constant temperature, so that the temperature in the high-density region could be obtained from the temperature in the region of infinitesimal densities. In this way, Fermi connected the quantum and the classical behaviour of the gas through a very effective and original application of the correspondence principle.

Fermi considered the ``$s$''-molecules simply as mass points, of energy $w=h\nu(s_1+s_2+s_3)=h\nu s$. ``The Pauli rule, therefore, states in our case that in the entire mass of gas at most only one molecule can have the given quantum numbers $s_1$, $s_2$, $s_3$.'' The cited equation admits $Q_s=(s+1)(s+2)/2$ solutions for a given $s$. Thus, according to the `generalised' Pauli principle, there exist at most $N_s \leq Q_s$ ``$s$''-molecules, otherwise two molecules would assume the same set of values $(s_1, s_2, s_3)$.\footnote{
For instance, for $s=1$ the values for $(s_1, s_2, s_3)$ may be (1,0,0), (0,1,0), or (0,0,1). One therefore has $Q_1=3$, and $N_1\leq3$.}
By arranging the $N_s$ molecules into the $Q_s$ places, a new statistical method was introduced, in addition to those of Boltzmann and of Bose-Einstein.

Fermi calculated the number $P$ of arrangements of the $N$ molecules, for which $N_0$ molecules were at places with energy 0, $N_1$ at places with energy $h\nu$, $N_2$ at places with energy 2$h\nu$, etc. By seeking the maximum of $P$ under the constraints $\sum N_s=N$ and $\sum sN_s=E$ (where $E=W/h\nu$), he obtained the distribution law: 
$$N_s=Q_s\frac{\alpha e^{-\beta s}}{1+\alpha e^{-\beta s}}=Q_s\frac{1}{\frac{1}{\alpha}e^{\beta s}+1} .$$
The values of $\alpha$ and $\beta$ could be obtained from the previous constraints; conversely, $\alpha$ and $\beta$ could be considered as given, and the total number and total energy of the configuration derived. 

Taking into account the expression for $N_s$, he calculated the density of the molecules with kinetic energy between $L$ and $L+dL$ at the position $x, y, z$. By comparing that density with the Maxwellian distribution, he found $\beta=h\nu/kT$. Instead, $\alpha$ could be defined only implicitly.

Fermi invoked the virial theorem to obtain the equation of state. In the limit of weak degeneration, that equation reduces to the classical equation of state of ideal gases, while the entropy converges to the Sackur-Tetrode formula. In the limit of large degeneration, he calculated the pressure and the mean kinetic energy, as well as the specific heat at constant volume, vanishing linearly with temperature.

Fermi's paper was published in one of the most prestigious international journals, but it can be said to have had no immediate impact on the international community; no more than the preceding Italian paper. Actually, it was still based on the Sommerfeld quantum condition and would appear markedly old fashioned. Dirac himself read Fermi's paper between May and August 1926, and later stated: ``I did not see how it could be important for any of the basic problems of quantum theory; it was so much a detached piece of work''.\footnote{
See \cite{DiracRecollections}, p. 133. Dirac also stated: ``The statistics [...] was first proposed by Fermi, and my later work showed how it could be fitted in with quantum mechanics,'' ibid, p. 134.}

In any case, Fermi gradually earned high reputation among the international community. Thus, in a letter to Lorentz of June 1927, Einstein proposed Fermi as his own substitute for giving a report on quantum statistics.\footnote{
Letter of A. Einstein to H.A. Lorentz, of June 17, 1927; reprinted in 
\cite{PaisEinstein}, p. 432. We recall that Fermi met Einstein in Leiden in September 1924.}
Fermi gained great acknowledgment and satisfaction from the Conference held in Como (and in Pavia and Rome) in September 1927, when Sommerfeld gave a talk entitled {\it Zur Elektronentheorie der Metalle und des Volta-Effektes nach der Fermi'schen Statistik} (On the electron theory of metals and of the Volta effect according to Fermi's statistics).\footnote{
\cite{Sommerfeld1}. Sommerfeld's contribution to the Proceedings of the conference was followed by
\cite{Sommerfeld2}.}
Sommerfeld stressed that many experimental results (e.g. the Wiedemann-Franz law, the Thomson effect, etc.), contrasting with the previous theory, could be explained on the basis of Fermi's statistics. As is well known, he gave also account of the contribution of the electrons to the specific heat of metals.
In the discussion following Sommerfeld's talk, Fermi suggested a research programme, aimed at developing the electron theory of metals.\footnote{
\cite{Como}, vol. II, pp. 470-71, and pp. 594-96. Fermi's discussions are also reported in \cite{FNM}, vol. I, pp. 179-81.}

Fermi applied his statistical method to many electrons atoms, for the first time, in a work entitled {\it Un metodo statistico per la determinazione di alcune propriet\`a dell'atomo} (A statistical method to evaluate some properties of the atom), which Corbino communicated to the Accademia dei Lincei in December 1927 (followed by another work in January).\footnote{
\cite{FermiAtomo1}, \cite{FermiAtomo2}.}
He treated the electrons as a completely degenerate gas surrounding the atomic nucleus, the average potential for each electron obeying a nonlinear differential equation that Fermi solved numerically. 

L.H. Thomas had already reached essentially identical conclusions one year before --- though Fermi was not aware of it, apparently --- and for this reason this statistical model was designated as the Thomas-Fermi model.\footnote{
\cite{Thomas}.}
It proved very useful in evaluating the electronic properties of heavy atoms, in fair agreement with the experimental results.

\section{The Fermi lectures on atomic physics in his first course in Theoretical Physics}
\label{sec:foo}

There existed no chair of Theoretical Physics at the Italian universities until mid-1920s. Corbino promoted the institution of such a chair, with the support from the very influential mathematicians of the University of Rome (among whom Tullio Levi-Civita).\footnote{
On December 22, 1924 the Faculty of Sciences at the Royal University of Rome agreed to the institution of a chair of Theoretical Physics, which was formally claimed on April 27, 1925.}
In November 1926, the board of examiners, composed of Corbino, M. Cantone, A. Garbasso, G.A. Maggi, and Q. Majorana, selected Fermi, Enrico Persico, and Aldo Pontremoli as suitable for a professorship. Fermi was appointed as a professor of Theoretical Physics at the Royal Institute of Physics in Rome, Persico in Florence, and Pontremoli in Milan.

The examiners exalted Fermi's attitude to physics: ``While he is fully able to handle the most subtle questions in Mathematics, he can use them soberly and discretely, without ever loosing sight of the physical problem he is trying to solve, and of the role and value of the physical quantities at hand''.\footnote{
The minutes of the meeting were published in \cite{Ministero}. The verdict of the board of examiners, about Fermi, Persico, and Pontremoli, is reported in
Ref. \ref{ftn:AnniItaliani}, pp. 172-73.}
The same attitude would characterise also his teaching approach ever after.

On January 20, 1927 Fermi gave his first lecture on Theoretical Physics in Rome. In 1955 Persico wrote that ``Fermi was a born teacher and to him teaching came naturally.'' Moreover: ``Fermi's teaching method, either direct or indirect, and his very personalised working style, raised the level, in a little more than 10 years, of the School of Italian Physics to a point which had previously been deemed impossible''.\footnote{
\cite{Persico}, p. 38.}
In the same year Edoardo Amaldi, who had attended Fermi's course in theoretical physics together with Ettore Majorana and Segr\`e in 1927-28,\footnote{
See \cite{DGE}.}
commemorated Fermi as a teacher as follows: ``Fermi arrived in Rome in 1926 [...]. He had assembled a small group of young people who were enthusiastic about physics and the new horizons that were opening in the field, and dedicated himself to their preparation. This involved [...] lessons in theoretical physics [...], which he taught with diligence and exemplary simplicity, presenting only what was essential, stripping the subject of any useless adjunctions''.\footnote{
\cite{AmaldiCommemorazione}, pp. 26-27.}

Amaldi stressed also that incisive clarity, strict reasoning, and stimulating criticism always marked Fermi's lectures. Hans Bethe, who had been reading in Rome in the early 1930s, further recalled the ``enlightening simplicity''\footnote{
See \cite{Bethe1}, p. 253, and \cite{Bethe2}, pp. 28-29.}
of Fermi's teaching. 

The topics of Fermi's 1926-27 lectures on theoretical physics are listed on a record book that is now kept at the Archives of ``Sapienza'' University of Rome.\footnote{
The teacher's record-books listing the topics Fermi lectured on in his courses --- in Theoretical Physics, but also in Mathematical and in Earth Physics --- are kept at the Archives of the ``Sapienza'' University of Rome; they are labelled on the basis of the Faculty, the Professor, the title of the course, and the academic year. The front-page of the teacher's record book reported: ``Mod. n. 10 / R. Universit\`a degli studi di Roma / Facolt\`a di Scienze matematiche, fisiche e naturali / LIBRETTO / delle lezioni di .......... / dettate dal sig. prof. .......... / nell'anno scolastico ..........'' The boxes where the topics of each lecture were recorded reported: ``Argomento della lezione ... / Addi ...192.../ Firma dell'insegnante.'' The record books of Fermi's 1928-29, 1929-30, and 1938-39 courses in Theoretical Physics are not in these Archives; that of 1929-30 is kept at the Domus Galilaeana in Pisa, and that of 1938-39 --- recording only three lectures, Fermi gave before leaving for the USA on December 6, 1938 --- is at the Archivio Amaldi, at the Physics Department of the ``Sapienza'' University of Rome. Fermi's 1926-27 and 1927-28 record books are reproduced at the following URL:  \url{www.phys.uniroma1.it/DipWeb/Fermi_Majo/FMajo.html}; their contents are reported and discussed in \cite{DGS-Debutto}.
}
Detailed pieces of information on the full contents of his lectures come from some lithographed notes, taken by Carlo Dei and Leonardo Martinozzi, consisting of 11 chapters, and 157 pages.\footnote{
The title-page of Fermi's lithographed notes reports: {\it Lezioni di Fisica teorica dettate dal prof. E. Fermi, Raccolte dai dott. Dei e Martinozzi, Anno scolastico 1926-27, Roma}. Dr. Leonardo Martinozzi was Corbino's assistant until 1927. Then he moved to the Presages Bureau [{\it Ufficio presagi}] of the Meteorological Office (directed by Filippo Eredia, who had been Fermi's teacher of Physics at the high school). Carlo Dei graduated in Mathematics at the Rome University in 1923, and in Physics in 1925. Between 1927 and 1930, he was frequently appointed as an examiner, for various University courses: in Experimental Physics, Advanced Physics, Terrestrial Physics, Theoretical Physics. We may guess he was one of the many high-school teachers cooperating in the University courses.}

Fermi expounded the most relevant results of atomic theory obtained within 1924. He accounted for them on the basis of the Einstein light quantum hypothesis (lectures 23-28), the Bohr atomic theory (29-38),
and its generalisation by Sommerfeld (39-43). He made no mention of the matrix mechanics (formulated in 1925), which was too poor of self-evident physical contents and used unusual mathematical tools. He made no mention of wave mechanics either: it had been introduced only few months before, and the meaning of the wave function was still lively debated among the physicists.\footnote{
See \cite{SegrKuhn}, p. 16. See also
\cite{PersicoRasetti-Kuhn}; in particular, as for Fermi's reception of the papers on matrix mechanics, Rasetti recalled that Fermi told him: ``Now I'm trying to read them and see what Heisenberg is trying to say, but so far I don't understand that'' (\cite{PersicoRasetti-Kuhn} p. 16).}
Thus, atomic physics, which represented the bulk of Fermi's course, was tackled within the framework of the old quantum physics. 

The course opened with the kinetic molecular study of the ideal gas, which represented one of the main arguments for the atomic theory of matter. The relation between the pressure and the mean kinetic energy per molecule was obtained without any use of differential or integral calculus: though theoretical physics was often mixed up with mathematical physics in Italy in those years, Fermi held that the understanding of the physical aspects was a top priority, and mathematics was just relegated to a secondary, operating role. Infinitesimal calculus and its methods were gradually introduced in the follow-up of his course. Fermi's purely phenomenological approach and his repeated reference to the results of experiments, whose essentials he described also through schematic diagrams, proved his course deeply different from those in mathematical physics. 

Fermi expounded the Maxwell electromagnetic theory, warning that it ``failed to account for the results that have in the last years come from the experiments'' on the interaction between electromagnetic radiation and matter. In particular, both the experimental features of the photoelectric effect and those of the Compton scattering of high-energy electromagnetic radiation (discovered in 1923) were accounted for by the Einstein light quantum hypothesis.\footnote{
Fermi dodged to stress that Planck (and later Bohr) for a long time held that a quantum of energy was continuously distributed in a spherical wave packet, whereas according to Einstein the energy was localised in an almost point-like region. Many handbooks also nowadays ignore such fundamental difference (see \cite{SebastianiDidattica}). As for the Compton effect, Fermi dealt with it without even mentioning relativity. To be precise, he neglected terms of order greater than the first in the difference $(\nu-\nu')$ of the light frequency, and at the same time he expressed the momentum of the electron simply as $mv$, thus neglecting the relativistic corrections: ``The curious affair is that an opposite error creeps in this way, so that the formula above [giving the wave-lenght shift] turns out to be strictly exact instead of being an approximation'' (pp. 81-82).}

Fermi considered the emission and absorption of radiation by atoms --- lecturing also on the experimental techniques proper to spectroscopy\footnote{
During his stay in Florence in the years 1924-26, Fermi collaborated with Rasetti and became familiar with experimental spectroscopy. Moreover, the early experimental activity of Fermi's group in Rome concerned exactly spectroscopy. 
See \cite{Messore}.}
--- and the Bohr theory of the atom.\footnote{
Differently from Bohr, who quantised the energy of the orbital electrons --- a condition easily transferred to the angular momentum --- , Fermi chose to quantise the orbital radius, a condition with a more immediate geometrical and kinematical meaning.}
He showed --- just as in his thesis --- that the latter accounted for  the characteristic X-rays spectra, ``with a good level of approximation.'' Further, he claimed an approximate application of the Bohr model to the many-electron atoms, whose spectra gave information about their electronic configurations.

Also the Sommerfeld conditions were considered in detail (note that Fermi had already used them to quantise the ideal gas, at the beginning of 1926). The exposition became more sophisticated when Fermi showed the invariance of the Sommerfeld conditions under any co-ordinates transformation. 

Fermi dealt extensively with the elliptic orbits of the hydrogen atom. Finally, he lectured on the Zeeman and the Stark-Lo Surdo effects, on the alkali spectra, on the Bohr correspondence principle, and on ``selection principle.'' 

The published notes follow very closely Fermi's 1926-27 lectures on theoretical physics, listed in his record book. Indeed, it is all the same whether we consider how many lectures, or how many pages in the notes were devoted to each topic. 

On the other hand, Fermi likely revised and improved the notes that Dei and Martinozzi had taken during his lectures. Actually, the Zeeman effect and the Stark effect are inverted in the published notes, compared to the order in which they were lectured on and listed in the record book. Furthermore, some passages of the published notes were repeated again in the {\it Introduzione alla fisica atomica}, which Fermi would publish in 1928.

The emphasis of Fermi's 1926-27 course was as follows:

\noindent Classical physics (42\%): 

\begin{itemize}

\item kinetic theory of gases (8\%); 

\item cathode rays (4\%); 
\item radioactivity (7\%); 
\item electromagnetic theory of light (23\%).
\end{itemize}

\noindent Quantum physics (58\%): 
\begin{itemize}
\item photoelectric effect (5\%); 
\item the Compton effect (7\%); 
\item the Bohr atom (7\%); 
\item spectral lines (10\%); 
\item the Sommerfeld conditions (13\%); 
\item electro- and magneto-electric effects (13\%); 
\item alkali spectra (3\%).
\end{itemize}

Classical physics extended for nearly a half of the course, the electromagnetic theory of light for about one quarter. The photoelectric and the Compton effects, the emission and absorption processes, and the line spectra, which challenged the wave theory of light, all covered about one third of the course; the Sommerfeld quantum conditions a little more than one eighth. 

A full discussion of various experiments and of their most relevant results was never missing, and gives us evidence of Fermi's phenomenological approach. We also recall that Fermi might intend such a plain, hardly formal approach as a way of stressing a priority for the physical contents; perhaps, also to mark his distance from those who he ironically called ``high priests of theory''.\footnote{
\cite{SegrTreccani}, p. 330 (\cite{D'Agostino}, p. 5).
}

\section{Fermi's second course in Theoretical Physics and his ``Introduzione alla fisica atomica'' }
\label{sec:foo}

Fermi's handbook {\it Introduzione alla fisica atomica}\footnote{
\cite{Introduzione}.}
was published by the publisher Zanichelli in autumn 1928. It was a broad adaptation of his 1927 lithographed {\it Lezioni}: it dealt with roughly the same topics as the notes did, but its extension was more than doubled. It consisted of 325 pages, and of 10 chapters. 

Chapter 1 (covering the 7\% of the total pages) deals with ``Kinetic theory of gases,'' in much greater detail than the lithographed notes. Chapter 2 (10\%) tackles the ``Electromagnetic theory of light,'' in a similar way to the notes. Chapter 3 (9\%) is devoted to the ``Electric corpuscles,'' and includes cathode rays, radioactivity, and the Rutherford model (three pages are now devoted to the ``Elements of relativistic dynamics''). Chapter 4 (8\%) deals with the ``Energy exchange between light and matter,'' i.e., with the light quanta and the Bohr frequency relation. This chapter introduces to the quantum physics, which covers one quarter of the book. 

Chapter 5 (22\%) is all devoted to ``The Bohr atom'' and the principles of the old quantum physics: the Sommerfeld conditions, correspondence principle, selection rules, the Ehrenfest adiabatic principle, spatial quantisation, the Bohr magneton, the spinning electron. Chapter 6 (17\%), ``Spectral lines'' [{\it Molteplicit\`a spettrali}], is a review of quantum spectroscopy, inspired by Sommerfeld's handbook, which had introduced Fermi to the foundations of atomic physics when he was in Pisa. Chapter 7 (6\%) deals with the ``R\"ontgen-rays spectra,'' a topic dear to Fermi, who had tackled it in his thesis. Chapter 8 (3\%) goes through ``Band spectra'' and molecular spectroscopy, which Fermi would broaden considerably in {\it Molecole e cristalli} six years later.\footnote{
\cite{Molecole}.}

Chapters 9-10 introduce some now topics, compared to the lithographed notes. Chapter 9 (10\%) deals with ``The statistics of the quantum theory'': specific heat of solids, black-body spectrum, entropy constant, the theory of the thermoelectric effect and that of paramagnetism, are treated within the Boltzmann distribution law. Incidentally, Fermi would introduce the Bose and the Fermi statistics only in {\it Molecole e cristalli}. 

Chapter 10 (8\%) is entitled ``The new quantum mechanics.'' Both matrix mechanics and wave mechanics are mentioned, though Fermi enlarged only on the latter and on some of its applications (i.e., the harmonic oscillator and the hydrogen atom). 

The {\it Introduzione alla fisica atomica} filled in a considerable gap among the university handbooks in Italy, at the same time giving good evidence of Fermi's high didactical skills. According to Persico, Fermi's ``prodigious capacity of immediately spotting the essential element in everything and getting to the heart of it in the simplest way was his principal natural gift. [...] This can be evidenced in his lessons and writings''.\footnote{
\cite{Persico}, pp. 42-43.}

As for Fermi's handbook, Segr\`e wrote in his biography: ``Until 1928 there was no Italian book on modern physics, suitable for training advanced university students. The text from which an entire generation had learned, Sommerfeld's {\it Atombau und Spektrallinien}, was in German; furthermore, it was too long and too detailed to serve as an introduction to the subject. Typically, Fermi decided that the best thing to do was to write his own book, which he did during the summer vacation of 1927 [... . This book] performed its task admirably. Unfortunately, it was written before wave mechanics had really taken hold, and for this reason, although Schr\"odinger's equation is treated in a chapter at the end of the book and Heisenberg's matrices are mentioned, the development of the subject is primarily based on the old Bohr-Sommerfeld orbit conception''.\footnote{
\cite{EFPhysicist}, p. 47. Segr\`e said that Fermi wrote his book in Val Gardena, in Summer 1927. However, Segr\`e probably referred here to the revision, by Fermi, of the notes taken by Dei and Martinozzi. Segr\`e's outline of the university handbooks in Italy should in part be re-shaped: the handbook of Leo Graetz, {\it Le nuove teorie atomiche e la costituzione della materia} (The New Theories of the Atom and the Constitution of Matter, \cite{Graetz}), in Italian, was already available when Fermi published his {\it Introduzione}. Graetz went through the kinetic theory of gases, the electron, the radioactive disintegrations, the Rutherford model of the atom, X-rays spectra, the Bohr theory of the hydrogen atom, elements of nuclear constitution; still, no mention of the Sommerfeld conditions on the phase integrals. It was a translation, ``for the young chemists,'' of a German handbook: Graetz was professor at the Munich University.}

There was a fair correspondence between Fermi's {\it Introduzione} and his 1927-28 course in Theoretical Physics. If we compare closely the two, we realise that the topics of the lectures match quite well the Chapters 1 to 6 of the book. 

Also this course had a `slow' start. It opened with a good 15\% of lectures (9 over 59 of real exposition) on the kinetic theory of gases (including the proof of the equipartition of the energy). The same issue took only 9\% of the book (falling to 7\% if we also consider Chapters 7 to 10). Classic electromagnetism covered 17\% of the course (10 lectures), compared to 13\% of the book (10\% the last four chapters included). Thus, classical physics occupied one third of the course, but just a little more than one fifth of the (lectured) topics of the book. It is apparent that Fermi aimed to expound well-established topics, before taking his students into the intricacies of quantum theory. 

Fermi's course did not tackle such issues as R\"ontgen-rays spectra, band spectra, quantum statistics, and the new quantum mechanics, which conversely were dealt with in Chapters 7 to 10 and represented about one fourth of the book. To be precise, Fermi did not mention wave mechanics at all in his 1927-28 lectures, a gap that should be on purpose, and not due to lack of time (a good six lectures summarised the whole course). 

We may suppose that he did not welcome, or even ignored, wave mechanics, but that was not the case. Effectively, Persico reported that he himself, Fermi, and Aldo Pontremoli, in summer 1926, had stimulating discussions on Schr\"odinger's papers that had just marked the birth of wave mechanics: ``Fermi had already acquired a deep insight in their content''.\footnote{
See Persico's foreword, in \cite{FNM}, vol. I, p. 222.}
Furthermore, Fermi's works of the second half of 1926 disprove his lack of interest in wave mechanics: he used the Born wave theory of scattering; he reflected on the meaning of some physical quantities and of the adiabatic principle, within the framework of wave mechanics; he calculated the orbital magnetic moment of an electron in a central field, resorting to ``the Schr\"odinger expression for the average current density.'' In spring 1927, he tried to ``introduce in the framework of wave mechanics the radiation reaction on an emitting atom''.\footnote{
See
\cite{FermiWellenmechanik},
\cite{FermiPersicoAdiabatiche},
\cite{FermiMoments},
\cite{FermiEmissione}.}

Fermi's decision to leave out  wave (and matrix) mechanics from his course seems to agree with a point of view of Levi-Civita. The latter stated at the Como Conference that, despite the success of matrix and wave mechanics, it was not worth  giving up the ``hybrid approach'' to atomic theory, represented by the Sommerfeld rules. Indeed, the introduction of the quantum postulate into ordinary mechanics was ``undoubtedly attractive, corresponding to elementary and concrete forms of physical intuition, and above all being suitable to lead to quantum relationships in the simplest way with the usual procedures of analitical mechanics''.\footnote{
{\sc Levi-Civita} in \cite{Bernardini}, p. 87.}

Anyway, though not in the official morning-course, in 1927-28 Fermi taught the Schr\"odinger equation in seminars, he gave to a small circle in the afternoon: to Amaldi, Majorana, Segr\`e, Rasetti, and sometimes Corbino.\footnote{
See \cite{Amaldi-Kuhn}, p. 18. See also \cite{SegrKuhn}, p. 11.}

In the years to come, Fermi would lecture on Chapter 10.\footnote{
Fermi's 1929-30 record book --- kept at the Fermi Archives at the Domus Galilaeana in Pisa (Miscellanea, V 13) --- listed 11 lectures on wave mechanics. Fermi dealt with the Schr\"odinger equation already in his 1928-29 course (whose record book is not at the Archives of the ``Sapienza'' University, and is not mentioned in the inventory of the Domus Galilaeana either: see \cite{Leone2}). According to the archive records kept at the ``Sapienza'' University, Fermi asked a candidate about the Schr\"odinger equation during an oral exam in October 1929.}
However, he would still introduce quantum mechanics within the framework of old quantum theory, and from a phenomenological point of view: as a matter of fact, he would regard  the historic approach as the most recommended one for teaching quantum physics in his courses.

We now discuss the main contents of the last chapter of Fermi's {\it Introduzione}. He warned, at the beginning: ``The union of ordinary mechanics with the Sommerfeld quantum rules may only to a rough approximation account for atomic phenomena.'' Fermi added that the Heisenberg and the Schr\"odinger theories might, independently one from the other, remedy to this shortcoming and allow to investigate the atom.

As for matrix mechanics, he only very briefly mentioned that it was ``very intricate from the formal point of view, mainly because it resorts to unusual concepts and tools.'' Besides his doubts on the formalism, Fermi had already stated serious reservations in a letter to Persico, on September 23, 1925: ``My impression is that during the past few months there has not been much progress, in spite of the formal results on the zoology of spectroscopic terms achieved by Heisenberg. For my taste, they have began to exaggerate their tendency to give up understanding things''.\footnote{
\cite{EFPhysicist}, p. 209. See also \cite{PersicoRasetti-Kuhn}, p. 16/1.}

Though matrix mechanics and wave mechanics had been proven equivalent to each other by Schr\"odinger, they were inspired by ``different, not to say opposite, ideas.'' Fermi only expounded ``in brief the foundations of wave mechanics, perhaps less organic than the Heisenberg mechanics, but much more intuitive and formally plain.'' Starting from the analogy between classical mechanics and geometrical optics at the basis of the Schr\"odinger theory, he compared the behaviour of a light quantum and of a point-like mass subject to a potential. He then illustrated Schr\"odinger's path to wave mechanics, a theory that acknowledged ``most naturally the existence of discontinuous quantum states.''

Then, Fermi gave account for the ``scalar field $\psi$.'' Indeed, an electron should according to Schr\"odinger be described by a wave packet, and its velocity to the group velocity. However, the wave packet $\psi$ would always spread with time. This difficulty might be overcome through Born's hypothesis: considering the electron as a point-particle, ``the square of the amplitude of the vibration $\psi$ gives, for each point, the probability to find the electron in that point.'' This picture implied a renunciation of the causality principle.

The most relevant features of the eigenvalues and eigenfunctions of the Schr\"odinger differential equation were explored ``with mathematical language.'' In particular, the equation of the harmonic oscillator admitted finite, continuous, single-valued solutions in a full set of values of the coordinates, whereas, in the old quantum theory, discrete energy levels\footnote{
``[If we assign] an arbitrary value to $E$, [the Schr\"odinger equation] has, in general, no everywhere finite, continuous, and single-valued solution; it instead has, for some particular values of $E$, which exactly give the energy levels of the system,'' p. 309.}
sprang ``arbitrarily,'' from the application of the Sommerfeld conditions. 

Fermi also tackled the hydrogen atom, looking for the solutions of the Schr\"odinger equation when the potential was $U=-e^2/r$. Moreover, resorting to the time-dependent equation ``we realise that the Schr\"odinger theory in a very natural way implies the Bohr frequency relations.'' 

Fermi commented on Davisson and Germer's experiments on electron diffraction: ``Certainly, they very impressively prove a fundamental analogy between corpuscles and light. It will be the duty of tomorrow's physics to work out a theory that --- perhaps by replacing the ordinary concepts of physics and kinematics with a statistical approach --- accomplishes the reconciliation between the two groups of phenomena: those suggesting a wave picture, and those ascribing a point-like structure to light and matter.''

Fermi's notes and handbook, as well as all his writings, show his almost absolute lack of interest for any conceptual or epistemological aspect of quantum mechanics: aspects like these were only touched on --- for example in the previous quotation --- if not ignored at all by him. 

In particular, he made no mention at all of the uncertainty relations\footnote{
See \cite{Harris}, \cite{Cassidy}.}
in his {\it Introduzione}, despite having Heisenberg worked out it more than one year before, and Bohr mentioned it in Como.\footnote{
See \cite{BohrComo}.}
Nor can we find any mention of the uncertainty principle in the teacher's record books, until 1930-31.\footnote{
Fermi mentioned the uncertainty principle under the headword `Atomo,' he wrote for the \book{Enciclopedia Italiana di Scienze, Lettere ed Arti} (Istituto G. Treccani, vol. V, Roma 1930), pp. 243-52; Fermi's article is reprinted (with others) in
\cite{D'Agostino}, but not in
\cite{FNM}. Note that Segr\`e said that Fermi ``taught as little as possible of the uncertainty principle and all that,'' also in his courses at the Chicago University in the 1950s. That was for ``pedagogical reasons,'' according to Segr\`e (\cite{SegrKuhn}, p. 24).}

Persico vividly recalled: ``What I remember about the uncertainty principle is that I was much impressed by Heisenberg's paper and when I spoke to Fermi about it, I was surprised to find that he was not so enthusiastic about it. I had the impression that he did not think it was so important as I believed. Probably Fermi was not much interested in philosophical aspects of physics and that was too philosophical''.\footnote{
\cite{PersicoRasetti-Kuhn}, p. 1/2}

\section{Conclusions}
\label{sec:foo}
We gave an account of Fermi's work within the framework of old quantum physics. Fermi took his first steps towards theoretical physics autonomously, even becoming ``a propagandist of quantum theory,'' while most of Italian physicists were foreign to the progress taking place in this field abroad.  He always showed a special preference for a phenomenological approach, never loosing sight of the physical problem he was dealing with. 

As we have seen, Fermi's work on quantum statistics was an exemplary combination of quantum rules and intuitive pictures: the exclusion, the adiabatic, and the correspondence principles, together with some very simple assumptions on the behaviour of the gas molecules, led Fermi to the new statistical methods, which he proved to be so fruitful also in the study of the many-electrons atoms.

The elucidation of the physical aspects of a problem -- through intuitive picture, if possible -- was a top priority for Fermi, all the more so in his teaching activity. This emerged clearly from his first two courses in Theoretical Physics. Their content can be known in detail, through a combined analysis of some archive documents, on the one hand, and of some published notes and Fermi's 1928 handbook, on the other hand. 

In his 1926-27 course, Fermi made no mention of matrix mechanics, which was too poor of self-evident physical contents, nor of wave mechanics, the meaning of its wave function being still debated. Instead, a full discussion of various experiments, and of their most relevant results, was not missing; Fermi recognised old quantum theory, with its intuitive pictures, as an effective theoretical framework for interpreting these results.

The topics of Fermi's 1927-28 course in Theoretical Physics match quite well the Chapters 1 to 6 of his {\it Introduzione alla fisica atomica} -- which he published in 1928, and filled in a considerable gap among the university handbooks in Italy. Again, Fermi did not mention matrix mechanics in his course; he already had a deep insight in Schr\"odinger's paper, and yet he did not mention wave mechanics either. As a matter of fact, he agreed with a then widespread attitude in Italy, which considered the old quantum physics ``evocative, and in harmony with simple and definite pieces of intuition,'' and in this respect not worth being given up.

\section*{Acknowledgments}
We are very grateful to the Accademia nazionale delle scienze detta dei XL, for having put at our disposal the microfilmed copy of the interviews with Kuhn. We also thank Dr. Antonella Grandolini, for her kind help in the consultation of the Archive of the Accademia.
We are very grateful to Prof. Roberto Vergara Caffarelli, of the Pisa University, and to the Director of the Library, for permission to consult a copy of Fermi's thesis.
The kind availability of Dr. Franza Azzaro and of Mrs. Rossana Nardella, as well of Mr. Enrico Amici, Mr. Angelo Jona, and Mr. Piero Lucidi, of the Archives of the ``Sapienza'' University of Rome, is gratefully acknowledged.

\end{document}